RESEARCH

# MCA: Multiresolution Correlation Analysis, a graphical tool for subpopulation identification in single-cell gene expression data

Justin Feigelman[1,2], Fabian J. Theis[1,2] and Carsten Marr[1*]

## Abstract

**Background:** Biological data often originate from samples containing mixtures of subpopulations, corresponding e.g. to distinct cellular phenotypes. However, identification of distinct subpopulations may be difficult if biological measurements yield distributions that are not easily separable.

**Results:** We present Multiresolution Correlation Analysis (MCA), a method for visually identifying subpopulations based on the local pairwise correlation between covariates, without needing to define an a priori interaction scale. We demonstrate that MCA facilitates the identification of differentially regulated subpopulations in simulated data from a small gene regulatory network, followed by application to previously published single-cell qPCR data from mouse embryonic stem cells. We show that MCA recovers previously identified subpopulations, provides additional insight into the underlying correlation structure, reveals potentially spurious compartmentalizations, and provides insight into novel subpopulations.

**Conclusions:** MCA is a useful method for the identification of subpopulations in low-dimensional expression data, as emerging from qPCR or FACS measurements. With MCA it is possible to investigate the robustness of covariate correlations with respect subpopulations, graphically identify outliers, and identify factors contributing to differential regulation between pairs of covariates. MCA thus provides a framework for investigation of expression correlations for genes of interests and biological hypothesis generation.

## Background

Heterogeneity in cellular populations has been the focus of many recent publications in areas such as embryonic stem cells [1], induced pluripotency [2], transcriptomics [3], and metabolomics [4]. In biological experiments, data often originate from a mixture of qualitatively differing subpopulations corresponding to e.g. distinct phenotypes in assays of cellular populations. For example, whole blood samples contain a mixture of distinct cell lineages which can be identified based on the presence of lineage-specific cell surface markers [5]. Embryonic stem cells have also been shown to exhibit heterogeneous expression of pluripotency factors critical for the maintenance of pluripotency in culture [1][6]. Indeed, there is increasing evidence for the existence of cellular subpopulations with possible noise-induced transitions between phenotypic attractors [7]. Thus it is clear that traditional techniques, which provide only population averages, may fail to resolve the true population heterogeneity.

Technologies such as flow cytometry, single-cell qPCR, mass cytometry and time lapse fluorescent microscopy are uniquely positioned to answer questions regarding the makeup of cellular populations. Each is able to yield quantitative measurements of cellular state, i.e. mRNA expression or protein copy number, which may be representative of the underlying subpopulations.

If the subpopulations are not already known, various methods exist to attempt to learn them on the basis of the data distribution. Classical techniques such as clustering may be useful for subpopulation identification if the subpopulations are readily separable in terms of expression levels [8]. Alternatively, more sophisticated machine-learning based approaches such as mixture models, (fuzzy) k-means clustering, multilayer perceptrons, self organizing maps, support vector machines, regression trees, and many others have also been applied to subpopulation identification (see

*Correspondence: carsten.marr@helmholtz-muenchen.de
[1]Institute of Computational Biology, Helmholtz Zentrum München, Ingolstädter Landstrasse 1, 85674 Neuherberg, Germany
Full list of author information is available at the end of the article

Lugli *et al.* [9] and Bashashati *et al.* [10] for a review of subpopulation identification approaches applied to flow cytometry).

However, existing methods for subpopulation identification predominantly rely on heterogenous expression levels. If the distributions overlap, identification of individual subpopulations based on expression alone may be difficult. In the case where subpopulations exhibit differential regulation motifs, they may be identifiable based on their distinctive correlations. Examining the local, state-dependent correlation of covariates provides additional information regarding the underlying distributions attributable to distinct subpopulations. In particular, we expect correlations to change for regions of state space (i.e. the space of possible gene expression levels) containing predominantly samples from a single subpopulation. Correlation analysis in subspaces of high dimensional data have gained attention over the past several years, particularly in the context of data mining e.g. in databases. For instance, algorithms such as MAFIA [11], CURLER [12], $\delta$-Clusters [13] , ENCLUS [14], etc. have been proposed for automatic identification of clusters using lower-dimensional subspaces. However, automatically identified clusters may be difficult to interpret biologically, and it may be difficult to assess their relative robustness.

We introduce a complementary method, Multiresolution Correlation Analysis (MCA) for systematically examining the dependence of local correlation upon location in state space. Using MCA, the correlations of pairs of variables are examined for regions of state space subdivided with varying granularity. The analysis can be summarized using MCA plots, which provide a visual representation of the pairwise correlation as a function of expression of a third variable.

MCA plots simultaneously visualize the correlations of data subsets of all sizes, centered at all locations in the distribution of a sorting variable, making it possible to distinguish regions with robust correlations which may be indicative of distinct subpopulations. Lastly, they provide the ability to identify observations which contribute disproportionately to the overall correlation structure, and hence skew the estimated correlation of the entire population.

## Results

### MCA reveals differential regulation of subpopulations in simulated gene expression data

To evaluate the MCA approach, we simulated gene expression data using a simple three species gene regulatory motif, given by Equation (6) as described in *Methods*. In this system, $Z$ activates $X$ and $X$ activates $Y$ (Figure 1A, left) via Hill-type activation functions, and population-level heterogeneity is introduced via the use of stochastic differential equations which approximate the intrinsic noisiness of gene expression [15][16][17].

The steady state distribution resulting from a typical simulation (Figure 1A, center) shows a significant positive Pearson correlation ($p < 0.05$) between $Z$ and $X$, and between $X$ and $Y$ (Figure 1A, right), and no significant correlation between $Z$ and $Y$, as would be expected from the underlying regulatory motif.

Similarly, we simulated a biological system for which $Z$ activates $X$, but where $X$ inhibits $Y$ (see Figure 1B, left) and Equation (7) of *Methods*). The resulting steady state distribution (Figure 1B, center) appears similar to that of the activation model. However, correlation analysis reveals that $Z$ and $X$ show significant positive correlation, and $X$ and $Y$ significant negative correlation (Figure 1B, right), in accordance with the underlying biological motif. The Pearson correlation also indicates significant negative correlation between $Y$ and $Z$ in the inhibition model, an indirect effect.

When combining the steady state distributions from activation and inhibition models (Figure 1C), the net Pearson correlation between $X$ and $Y$ is significantly negative (Figure 1C, I). Absent of subpopulation analysis, we would conclude that the relationship between expression levels of $X$ and $Y$ is antagonistic, implying an inhibitory motif.

In contrast, performing the same analysis on the subpopulation with $Z$ expression levels in the lowest 30% of the $Z$-distribution (Figure 1C, II) yields a significant positive correlation between $X$ and $Y$. Likewise, performing correlation analysis on the samples in the top 30% of the $Z$-distribution shows just the opposite, a significant negative correlation between $X$ and $Y$ (Figure 1C, III).

We can combine all of the $Z$-sorted subpopulations of varying size together using the MCA plot (Figure 1D), constructed as described in *Methods.* Briefly, the MCA plot shows the correlation of a pair of factors, for subpopulations defined by a sorting variable. The abscissa indicates the median value of the sorting variable for that subpopulation and the ordinate indicates the fraction of the population included in that subpopulation. Thus, higher points indicate larger subpopulations, points to the left indicate lower overall expression of the sorting variable, points to the right higher overall expression, etc. The regions where the computed correlation is statistically significant ($p < 0.05$) are indicated.

By systematic inspection via the MCA plot, we can conclude that subpopulations with low $Z$ values indeed show significant positive correlation between $X$ and $Y$ (Figure 1D, blue region), and subpopulations with high $Z$ values show significant negative correlation between $X$ and $Y$ (Figure 1D, red region, see *Methods* for details).

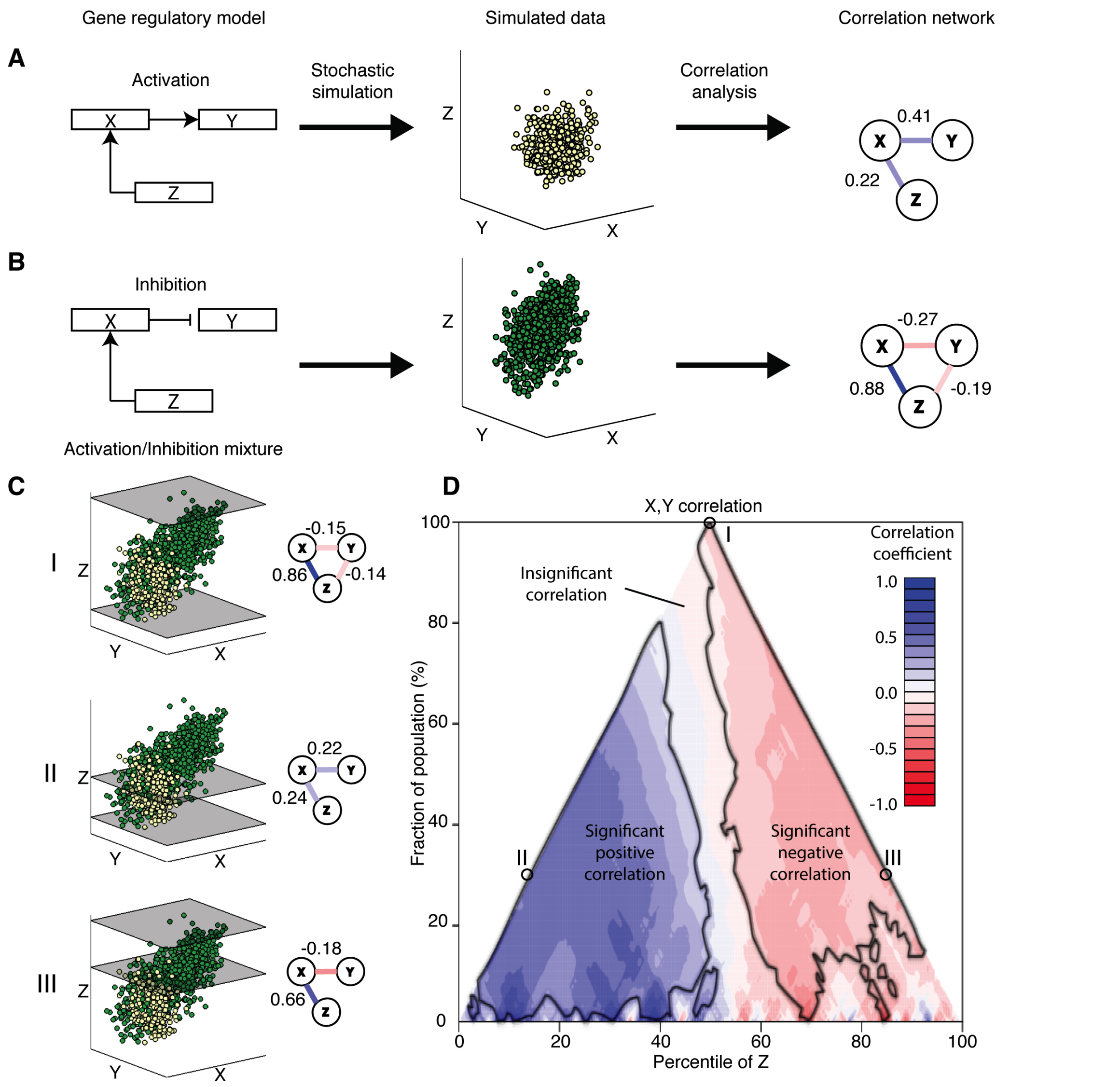

**Figure 1 MCA reveals the presence of subpopulations with differential regulation. A.** A three species activation motif (left), its steady state distribution (center) from an SDE simulation, and the resultant correlation network (right), showing positive correlation for species with an activating interaction. **B** A three species activation/inhibition motif induces positive correlation corresponding to activation and negative correlation corresponding to inhibition. **C.** Mixture of the activation and inhibition steady state data depicted in A and B. **I. II.** Correlation analysis of the subset from the lowest 30% of the $Z$-distribution shows significant positive $X, Y$ correlation. **III.** Correlation analysis of the subset from the highest 30% of the $Z$-distribution shows significant negative $X, Y$ correlation. **D.** Combining all subpopulations sorted by median $Z$ value and subpopulation size into an MCA plot reveals robust separation of positive and negative correlations for subpopulations with low or high $Z$ values, respectively.

### MCA plots as a diagnostic tool for transcriptomic analysis

MCA plots can be used to provide a multiresolution view of the correlation structure of real transcriptomic data. This allows us to confirm previous conclusions regarding heterogeneous subpopulations, detect potential novel subpopulations, and provides insight into the origin of the observed correlations.

We used MCA to analyze previously published single-cell transcriptomic data obtained from mouse embryonic stem cells (mESCs) [18][19]. There, microfluidic single-cell qPCR was used to obtain the relative expression of mRNAs for eight transcription factors known to be involved in regulation of pluripotency in mESCs: Fgf5, Nanog, Oct4, Sox2, Rex1, Pecam1, Stella and Gbx2, and Gapdh, a housekeeping gene against which all other transcript copy numbers were normalized. Analysis of subpopulations showed difference in the correlation networks of Nanog+/- and Fgf5+/- subpopulations, as well as clear separation of subpopulations using principal component analysis.

After data cleaning and normalization according to the method of Trott *et al.* [18], we generated the MCA plots for all pairs of genes, for all possible sortings, using Pearson correlation and a significance cutoff of $p < 0.05$. All points with $p > 0.05$ are colored white in the MCA plot.

*Detection of robust correlations* In an MCA plot, correlations that are globally robust with respect to changes in the sorting variable are easily distinguished by uniform coloration. For example, the correlation of Rex1 and Sox2 is robust with respect to changes in Pecam1 expression (Figure 2A, top). The scatter plot of Rex1 and Sox2 is shown for reference (Figure 2A, bottom). The robust positive correlation of Rex1 and Sox2 is consistent with current models of transactivation of Sox2 by Rex1 [20].

*Outlier detection* Correlation analysis can be sensitive to one or a few samples which substantially alter the estimated correlation of the entire population. In such a case, all subpopulations including these samples show a significant correlation, whereas their exclusion results in no significant correlation or potentially correlation of the opposite sign. MCA plots are able to detect such samples and identify them as sources of the detected correlation. For example, when sorting by Sox2, all subpopulations which do not contain the sample with the highest Sox2 expression do not show statistically significant correlation between Rex1 and Gbx2, whereas all subpopulations that do include this point show significant positive correlation (Figure 2B, top). Upon inspection of the data (Figure 2B, bottom) it is obvious that this single point, indicated by the arrow, is an outlier. Exclusion of this point renders the Rex1, Gbx2 correlation insignificant.

*Subpopulation identification* MCA plots are useful for identification of interesting subpopulations as shown for synthetic data (Figure 1C). Regions exhibiting a robust correlation may indicate the presence of differential regulation or a distinct cellular phenotype. For instance, sorting by Stella reveals the presence of a large region (the highest 40% of the population) for which the correlation between Nanog and Oct4 is not statistically significant (Figure 2C, top). Conversely, including the cells from the lowest 60% of the Stella distribution is sufficient to induce a significant positive correlation (Figure 2C, top). Inspection of the scatter plot of Nanog and Oct4 (Figure 2C, bottom) confirms that the lower 60% is noticeably more correlated than the top 40%. Hayashi *et al.* [19] note that mESCs with low or absent Stella expression may be more representative of epiblast-derived stem cells, and thus are expected to show differential regulation from the high Stella cells, which are more embryonic stem cell-like. Interestingly, the possibility of antagonistic regulation between Oct4 and Nanog in mESCs has recently also been raised [21].

### MCA provides additional insight into previously described subpopulations

In order to identify subpopulations with different coexpression networks, Trott *et al.* [18] grouped cells according to normalized pluripotency gene expression. Networks are constructed on the basis of significant Pearson correlation between nodes, and subdivided into groups based on the presence of two heterogeneously expressed transcription factors, Nanog and Fgf5. The high Nanog (Nanog+) compartment was defined such that Fgf5 expression is absent for all cells with Nanog expression at or above the minimum level of this compartment.

*MCA plots confirm differential Gbx2, Sox2 correlation for high Nanog cells* As in their study, we find that the Nanog+ subpopulation indeed has a significant positive Pearson correlation between Gbx2 and Sox2 (Figure 3A, I). Also in agreement, the remaining cells (Nanog-, $0^{th} - 74^{th}$ percentile), show no significant correlation between Gbx2 and Sox2 (Figure 3A, II). However, we learn from the MCA plot that in fact only the top 10% contribute to the observed positive correlation; the subset of the high Nanog subpopulation between the $74^{th}$ and $93^{rd}$ percentile (Figure 3A, III) is not significantly correlated ($p = 0.57$).

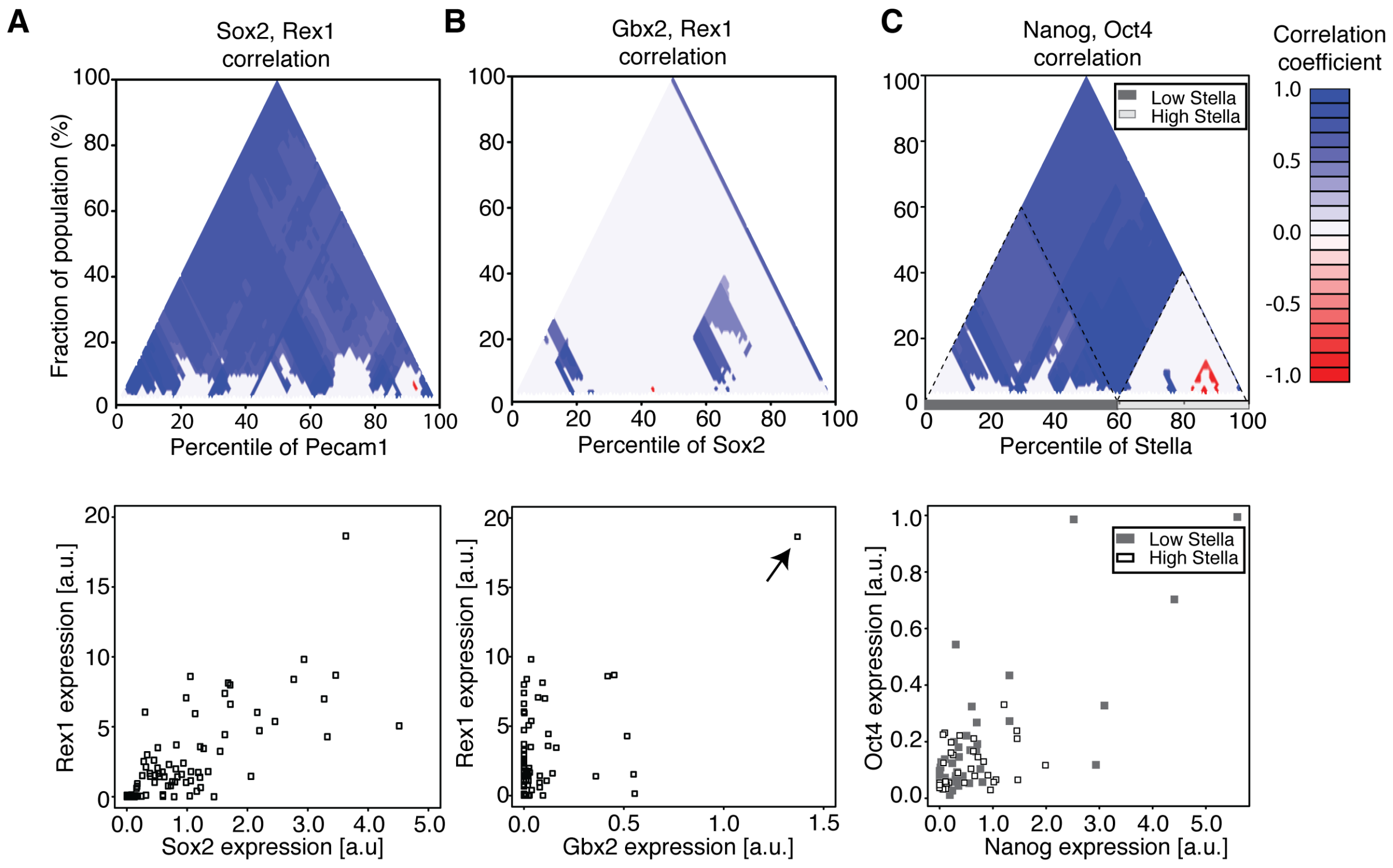

**Figure 2 MCA plots reveal important features of the correlation structure in single-cell transcriptomics data. A.** MCA plots with uniform appearance (top) reveal robust correlations amongst pairs of variables (scatter plot, bottom) like Rex1 and Sox2, sorted by Pecam1. **B.** Outliers can easily be detected via characteristic diagonal stripe patterns. Here a single sample with the highest value in the Sox2 distribution is enough to induce an overall positive Gbx2, Rex1 correlation (bottom, arrow). **C.** Robust subpopulations can be identified. The presence of a large triangular region with uniform correlation or lack of correlation between Rex1 and Nanog may indicate a subpopulation, seen here for cells from the highest 40% of the Stella distribution (top). The cells from the high Stella compartment (open boxes) are not significantly correlated for Rex1 and Nanog, in contrast to those from the low Stella compartment (filled boxes, bottom).

*MCA plots show that Gbx2, Sox2 correlations are not robust for Fgf5- cells* The authors found that the 15 of 83 cells (18%) expressing Fgf5 (Fgf5+ compartment) do not correlate for Gbx2 and Sox2, whereas the remaining 68 Fgf5- cells (82%) show a significant positive correlation [18] . Using an MCA plot we see that this indeed true (Figure 3B, I and II for Fgf5+, Fgf5-, respectively). However it is also evident that the Fgf5+ cells with Fgf5 expression between the $90^{th}$ and $100^{th}$ percentile of the distribution are in fact positively correlated for Gbx2 and Sox2 (Figure 3B, III). Likewise, the majority of the cells in the Fgf5- compartment are not significantly correlated for Gbx2 and Sox2. Indeed most subpopulations consisting of cells with expression between the $0^{th}$ and $75^{th}$ percentile of the Fgf5 distribution are not significantly correlated for Gbx2 and Sox2 ($p > 0.05$). Thus, MCA provides the means for a detailed and robust subpopulation identification, superior to *ad hoc* compartmentalization.

## Discussion

Fueled by newly developed single-cell technologies such as single-cell transcriptomic [22][23], genomic [24] and proteomic [25] analysis, many new methods have emerged which attempt to shed light on cellular heterogeneity [26][27][28][29].

Previous methods for the detection of heterogeneous subpopulations in biological data have largely focused on grouping observations according to expression level, and thus requires that subpopulations be readily separable. For instance, in FACS cellular subpopulations are often identified with manually determined compartments [30][31][32]. If the data are easily separated, clustering methods such as Gaussian mixture modeling and k-means clustering have proven well suited to this task [8].

Alternatively, methods such as principal component analysis attempts to identify the principal directions, along which the data are maximally separated [33]. Data which cluster together in the reduced dimensional subspace spanned by the first few principal components are thought to be representative of subpopulations. A similar method was employed by Trott *et al.* when analyzing the Fgf5+/- and Nanog +/- compartments [18]. Non-linear alternatives to PCA including Gaussian Process Latent Variable Modeling have also recently been shown to be useful for the identification of cellular subpopulations [29][34].

None of the previously mentioned methods utilize correlation information in the identification of cellular subpopulations, with the exception of Gaussian mixture modeling which attempts to learn the correlation matrices of Gaussian distributions thought to have generated the data. However, as shown here, the local correlation structure provides additional insight into the existence of differentially regulated subpopulations and hence should not be disregarded.

To date, relatively few methods have addressed the possibility of local, state-dependent correlations. Chen *et al.* [35] developed a method for analyzing the effect of local non-linear correlations in gene expression data, and applied it to a microarray dataset; a similar method was recently developed by Tjøstheim *et al.* [36] for estimating local Gaussian correlation in the context of econometric data. However, these methods required the definition of a interaction scale for the computation of local correlations or consider only the relative distance between data points and not their absolute levels when computing local correlations.

Recently Cordeiro *et al.* [37], developed a sophisticated algorithm for identifying clusters of arbitrary orientation, also in a multiresolution context. MCA is not as general in that it does not consider clusters aligned along arbitrary projections of the data but provides instead a comprehensive, multiresolution view of the correlation structure according to the measured covariates, preserving expression-level dependencies while not requiring any predefined bandwidth or interaction distance, and thus may provide more biological insight into the role of individual factors in differential regulation motifs.

MCA has the advantage of being easy to compute and intuitively interpretable; it is in effect a moving window correlation analysis simultaneously over many window sizes. The MCA plot provides a graphical diagnostic for detection of subpopulations points that contribute inordinately to the overall correlation, or outliers, and may provide biological insights that serve as hypotheses for further experimentation. Finally, although we have focused on biological data and in particular cellular subpopulations in single-cell transcriptional data, the method is more general and applicable to any multivariate data.

While the simplicity of MCA plots makes them easy to interpret, there are nonetheless shortcomings that must be mentioned. MCA plots are a graphical representation of the interaction of only two factors, sorted by a third. If there are many covariates, many such plots are possible, and it becomes increasingly more difficult to generate and search through all possible plots as the dimension increases. In such cases it is helpful to consider only those plots which may be of biological interest such as sorting variables thought to have a regulatory role, or pairs of factors that are suspected to interact. However, one may also use alter-native sorting variables, such as products of covariates representing potential interactions, principal directions as determined by PCA, or even arbitrary nonlinear functions of the covariates.

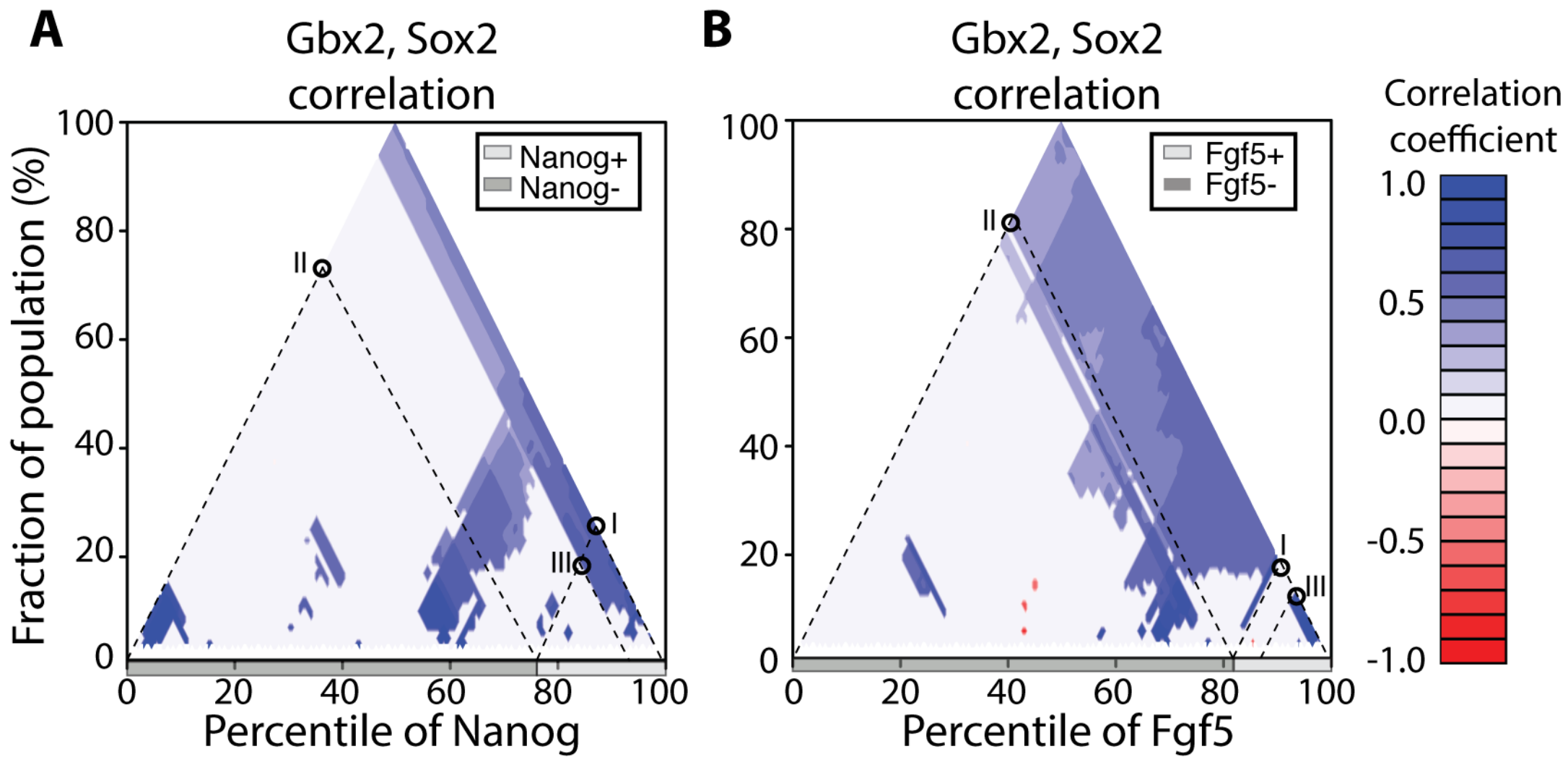

**Figure 3 MCA plots identify interesting biological subpopulations in mouse embryonic stem cells. A.** MCA analysis reveals insight into the influence of Nanog on the Gbx2, Sox2 interaction. Gbx2 and Sox2 are significantly positively correlated when considering the entire Nanog+ compartment (quantiles 74% to 100% of Nanog, **I**). When considering the remaining Nanog- cells, the correlation is no longer significant (quantiles 0% to 74%, **II**). MCA plots reveal that the positive correlation in the Nanog+ compartment is due to just half of the compartment; the rest is uncorrelated (**III**). **B.** Gbx2 and Sox2 are uncorrelated when considering the whole Fgf5+ compartment (quantiles 82% to 100%, **I**). However, the top 10% are significantly positively correlated when considered alone (**II**). The Fgf5- compartment is significantly positively correlated (**III**), however, the majority of subpopulations in the Fgf5- compartment are not significantly correlated. See main text for a comparison of our findings with the previous report of Trott *et al.* [18].

In the case of many variables, one may wish to sort the resultant plots according to arbitrary functions of the estimated correlation structures; i.e. one could filter for only those plots showing large significant regions or for plots for which a significant region of both positive and negative correlation are present. Although preliminary tests with such methods are successful in identifying such interesting plots, the results are not shown here as they are unnecessary when the number of dimensions is still manageable via manual inspection.

The correlation becomes difficult to estimate when the number of samples is small, or when the number of variables is relatively large compared to the number of observations. If the resolution is fine, then the MCA plot will contain many points for which the corresponding subpopulation only contains one or a few observations. Such points are omitted from the plot since the correlation cannot be robustly computed. This can sometimes give rise to small regions near the bottom of the MCA plots for which there are too few observations to compute the subpopulation correlation. These regions do not have biological significance.

Similarly, the stochastic nature of the data may give rise to ”noise” in small subpopulations, leading to interspersed points on the MCA plot which are not part of a large, significant region. These points typically do not indicate robust subpopulations since a small perturbation away from them leads to a different correlation structure, and can safely be ignored. This ”noise” also gives rise to the slight inhomogeneities in the regions identified in Figures 2 and 3.

Lastly, in the case of relatively many variables compared to the number of observations, correlations can be computed using shrinkage-based estimators [38], although this results in a different estimation of statistical significance, and increases computational complexity.

## Conclusion

We have presented a method for the analysis of local correlation structures in subpopulations of multivariate data. MCA provides a multiresolution summary of correlations between pairs of variables as ordered by a third sorting variable. Using MCA, it is possible to detect robust correlations, identify outliers which can bias correlation estimates, and potentially discover new subpopulations or interactions giving rise to novel biological hypotheses.

Future work will focus on the development of methods to automatically identify variable pairs showing differential regulation in conjunction with a sorting variable, alleviating the need to manually search through plots for interesting behaviors.

## Methods

We introduce Multiresolution Correlation Analysis (MCA) as a means for visually analyzing the local correlation structure of pairs of covariates, sorted by a sorting variable.

### Estimation of correlations

The empirical estimation of the Pearson correlations of a pair of random variables is computed in the usual way, such that for a pair of random variables $X$, $Y$:

$$\widehat{cor}(X, Y) = \frac{1}{M-1} \sum_{i=1}^{M} \left(X_i - \bar{X}\right)\left(Y_i - \bar{Y}\right) \quad (1)$$

for a set of realizations $i = 1, \ldots, M$ of $X$ and $Y$.

If the data are not multivariate normally distributed, it is preferable to use a more robust measure of statistical correlation. For instance, Spearman's rank correlation coefficient is defined as in (1), but using the rank-transformed data [39]; it provides a non-parametric measure of correlation between a pair of covariates.

### Multiresolution correlation analysis

We define the matrix

$$D = \begin{bmatrix} d_{11} & \ldots & d_{1N} \\ \vdots & \vdots & \vdots \\ d_{1M} & \ldots & d_{MN} \end{bmatrix} = \begin{bmatrix} \vec{d_1} & \vec{d_2} & \ldots & \vec{d_N} \end{bmatrix}$$

as the matrix of observed data, where the rows correspond to individual observations, and columns to measured variables. Note that the data matrix is defined as the transpose of the data matrix employed in some other transcriptomic analysis methods.

Given $D$, we can compute the sample correlation between any pair of variables, for any subset of the total observations. In particular we examine subpopulations defined by different intervals within the distribution of $\vec{d_s}$, the $s^{th}$ column of $D$, for any desired sorting variable $s$. For example, we can examine subpopulations for which the value of $s$ is in the highest or lowest 30% of its distribution.

For a subpopulation centered on the $\alpha^{th}$ quantile of the sorting variable $\vec{d_s}$, and containing $\beta \times 100\%$, of the total observations, such that

$$\begin{aligned} 0 < \quad & \beta \quad \leq 1 \\ \frac{\beta}{2} < \quad & \alpha \quad < 1 - \frac{\beta}{2} \end{aligned} \quad (2)$$

we can compute the sample correlation matrix $\hat{\Sigma}(\alpha, \beta; s)$

$$\hat{\Sigma}(\alpha, \beta; s) = \{\hat{\sigma}_{ij}\}_{i,j=1\ldots N} \quad (3)$$

with

$$\hat{\sigma}_{ij} = \widehat{cor}(\vec{d_i}(\alpha, \beta; s), \vec{d_j}(\alpha, \beta; s)) \quad (4)$$

and

$$\vec{d_q}(\alpha, \beta; s) = \left\{ d_{pq} \middle| Q(\alpha - \beta; s) \leq d_{pq} \leq Q(\alpha + \beta; s) \right\} \quad (5)$$

where $Q$ is the quantile function, i.e. $Q(x; s)$ is the $x^{th}$ quantile of the distribution of $\vec{d_s}$, and $\vec{d_q}(\alpha, \beta; s)$ is the subset of the $q^{th}$ column of $D$ for which the sorting variable falls between the $(\alpha - \beta)^{th}$ and $(\alpha + \beta)^{th}$ quantile of its distribution.

We define $\Omega$ to be the set of all pairs $(\alpha, \beta)$ for which Equation (2) is satisfied; for all $(\alpha, \beta) \notin \Omega$, $\hat{\Sigma}(\alpha, \beta; s)$ is undefined. Intuitively, Equation (2) constrains $\alpha$ and $\beta$ such that the subpopulation can extend no lower than the minimum, and no higher than the maximum of the sorting variable.

Although any function could be computed for the subpopulations, we restrict ourselves to Pearson correlation. If there are relatively many variables compared to the number of observations, i.e. $N > M$, estimation of the correlation matrix becomes numerically infeasible. In this case, estimation of the correlation can be computed using shrinkage-based approaches such as implemented in the GeneNet R-package [38].

### Construction of MCA plots

We systematically investigate the correlation of the subpopulations defined by $(\alpha, \beta) \in \Omega$. This information can be condensed into a MCA plot for any pair of variables $(i, j)$ by plotting the magnitude of the $(i, j)^{th}$ entry of $\hat{\Sigma}(\alpha, \beta; s)$, with a color scale mapped to the interval $[-1, 1]$.

While $\hat{\Sigma}(\alpha, \beta; s)$ is in principle defined for all $(\alpha, \beta) \in \Omega$, in practice we choose $\beta = 1/R, \ldots, 0.5$ and $\alpha = \beta, \beta + 1/R, \ldots, 1 - \beta$ for some positive odd integer $R \leq M$ which determines the resolution of the MCA plot, i.e. the number of subpopulations examined: the larger $R$, the finer the resolution of the MCA plot.

For each computed subpopulation, a p-value is computed that depends on both subpopulation size and magnitude of the estimated correlation coefficient. Thresholding to retain only small p-values may reveal large subpopulations with strong correlations. However, due to the interdependence of the subpopulations (i.e. the estimated correlation coefficient of a subpopulation is determined by the correlation coefficients of the points below), it is not possibly to directly interpret the p-values as the probability of non-zero correlation.

Lastly, the number of possible MCA plots $N$ increases cubically with the number of variables $k$, i.e. $N = k(k-1)(k-2)/6$, rendering fully-automatic analysis difficult. In this case, it is recommended to consider sorting variables which are of potential biological interest, such as those that are known to be heterogeneously expressed.

## Implementation

MCA and the MCA plots were implemented using the `R` programming language. The routine allows the user to pass a data frame containing observations, select a sorting variable, and a subset of factors whose pairwise correlations are to be analyzed; choose color options, and the number of subpopulations (resolution); specify correlation method (Pearson, partial, or Spearman), enable significance cutoffs with user-specified p-value threshold, and optionally to save resulting plots. The algorithm works by iterating through all subpopulations defined by median quantile of the sorting variable and size of the subpopulation, and computing the corresponding correlations using the built-in routines for correlation and significance estimation. Code is available upon request.

## Stochastic simulation

Synthetic data were generated via simulation of a gene regulatory network, the dynamics of which obey a stochastic differential equation. Two cases were simulated: a three species activation model where of $Z$ activates $X$, and $X$ activates $Y$ (Figure 1A, top); and an inhibition model for which $Z$ activates $X$ and $X$ inhibits $Y$(Figure 1B, top).

The activation model obeys

$$\begin{aligned} \frac{dX}{dt} &= \frac{Z^{n_x}}{Z^{n_x} + {K_{zx}}^{n_x}} V_x - \beta_x \cdot X + \sigma \xi_X(t) \qquad (6) \\ \frac{dY}{dt} &= \frac{X^{n_y}}{X^{n_y} + {K_{xy}}^{n_y}} V_y - \beta_y \cdot Y + \sigma \xi_Y(t) \\ \frac{dZ}{dt} &= k_z - \beta_z \cdot Z + \sigma \xi_Z(t) \end{aligned}$$

and the inhibition model obeys

$$\begin{aligned} \frac{dX}{dt} &= \frac{Z^{n_x}}{Z^{n_x} + {K_{zx}}^{n_x}} V_x - \beta_x \cdot X + \sigma \xi_X(t) \qquad (7) \\ \frac{dY}{dt} &= \alpha_y + \frac{V_y}{X^{n_y} + {K_{xy}}^{n_y}} - \beta_y \cdot Y + \sigma \xi_Y(t) \\ \frac{dZ}{dt} &= k_z - \beta_z \cdot Z + \sigma \xi_Z(t) \end{aligned}$$

where model parameters are not necessarily the same between the activation and inhibition models.

In both cases, the drift of $X$ is a sigmoidal function of $Z$ and $Z$ is an unregulated birth-death process. Each species is subject to linear decay and stochasticity enters through the homogeneous Wiener processes $\xi_X(t), \xi_Y(t)$, and $\xi_Z(t)$ which are independent, with unit variance, and scaled by the factor $\sigma$.

The two systems were constructed in such a way that the steady state distributions do not fully overlap, but are instead displaced with respect to one another such that the inhibition model shows an approximately 40% increase in $X$, and 20% increase in $Z$ with respect to the activation model.

Parameters and initial conditions used for the activation model are given in Table 1, and in Table 2 for the inhibition model. Simulations were performed using a Euler-Maruyama SDE integration scheme [41] with time step $\Delta t = 0.1$, implemented in MATLAB. The resulting simulations were allowed to converge to the steady state distribution by discarding the first 300 data points, and subsequently thinned by a factor of 20. Pearson correlations were computed using the `corr` built-in function of MATLAB.

## Analysis of transcriptomic data

Single-cell transcriptomic data from 87 mouse embryonic stem cells were obtained from Trott, *et al.* [18] as an Excel spreadsheet containing qPCR readouts for eight pluripotency factors and one housekeeping gene. The expression of each gene was first adjusted by adding the minimum expression over all genes, 0.0217, and subsequently normalized by dividing by the expression of the gene *Gapdh* on a cell-wise basis.

Two cells were excluded due to the presence of missing data for some factors, and two additional cells were removed because they were thought to be outliers. The remaining 83 cells were subdivided into a Nanog+ compartment ($N = 20$), defined as the 20 cells with the highest Nanog expression, and for which no Fgf5 expression was detected, and the complementary Nanog- compartment ($N = 63$). The cells were separately divided into a Fgf5+ ($N = 15$) compartment, for which Fgf5 expression was detected, and a Fgf5- ($N = 68$) compartment with no Fgf5 expression.

Correlation networks were computed using Pearson correlation of the normalized data without any log transformation, and with a significance cutoff of 0.05.

**Competing interests**
The authors declare that they have no competing interests.

**Authors' contributions**
JF created the MCA plot, implemented the method, performed all analysis, created figures and wrote the manuscript. CM provided supervision with methods and commented on the manuscript. JF, CM and FJT designed the study.

**Authors' information**
Justin Feigelman is a doctoral candidate at the Technical University of Munich and the Helmholtz Zentrum München für Gesundheit und Umwelt (HMGU). Carsten Marr is group leader of Quantitative Single Cell Dynamics at HMGU. Fabian J. Theis is Director of the Institute of Computational Biology at HMGU and Professor at the Technical University of Munich.

**Acknowledgements**
We would like to acknowledge Michael Schwarzfischer, Michael Strasser, Jan Krumsiek, Philipp Angerer, Florian Buettner, Ivo Sbalzarini, Timm Schroeder and Adam Filipzyck for valuable discussions and feedback, and Jamie Trott and Alfonso Martinez-Arias for providing transcriptomic data. Funding was provided by the European Research Council (starting grant LatentCauses), and the Deutsche Forschungsgemeinschaft (SPP 1356 Pluripotency and Cellular Reprogramming).

**Author details**
[1]Institute of Computational Biology, Helmholtz Zentrum München, Ingolstädter Landstrasse 1, 85674 Neuherberg, Germany. [2] Department of Mathematics, Technical University of Munich, Arcisstrasse 21, 80333 München, Germany.

**Tables**

**Table 1** Model parameters used for activation model (Figure 1A).

| Parameter | Value | Description |
|---|---|---|
| $n_x$ | 2 | Hill coefficient of X activation |
| $n_y$ | 2 | Hill coefficient of Y activation |
| $K_{zx}$ | 900 | Equilibrium constant of X activation |
| $K_{xy}$ | 1000 | Equilibrium constant of Y activation |
| $V_x$ | 600 | Velocity of X production |
| $V_y$ | 600 | Velocity of Y production |
| $k_z$ | 450 | Basal production of Z |
| $\beta_x$ | 0.3 | Death rate of X |
| $\beta_y$ | 0.3 | Death rate of Y |
| $\beta_z$ | 0.5 | Death rate of Z |
| $X_0$ | 100 | Initial X |
| $Y_0$ | 100 | Initial Y |
| $Z_0$ | 100 | Initial Z |
| $\Delta t$ | 0.1 | Time step |

**Table 2** Model parameters used for inhibition model (Figure 1B).

| Parameter | Value | Description |
|---|---|---|
| $n_x$ | 2 | Hill coefficient of X activation |
| $n_y$ | 2 | Hill coefficient of Y activation |
| $K_{zx}$ | 4000 | Equilibrium constant of X activation |
| $K_{xy}$ | 1000 | Equilibrium constant of Y inhibition |
| $V_x$ | 10000 | Velocity of X production |
| $V_y$ | 70 | Velocity of Y production |
| $k_z$ | 110 | Basal production of Z |
| $a_y$ | 70 | Basal production of Y |
| $\beta_x$ | 0.5 | Death rate of X |
| $\beta_y$ | 0.1 | Death rate of Y |
| $\beta_z$ | 0.1 | Death rate of Z |
| $X_0$ | 100 | Initial X |
| $Y_0$ | 1500 | Initial Y |
| $Z_0$ | 1000 | Initial Z |
| $\Delta t$ | 0.1 | Time step |